\PassOptionsToPackage{unicode}{hyperref}
\PassOptionsToPackage{hyphens}{url}
\PassOptionsToPackage{dvipsnames,svgnames,x11names}{xcolor}
\documentclass[
  letterpaper,
  DIV=11,
  numbers=noendperiod]{scrartcl}

\usepackage{amsmath,amssymb}
\usepackage{iftex}
\ifPDFTeX
  \usepackage[T1]{fontenc}
  \usepackage[utf8]{inputenc}
  \usepackage{textcomp} 
\else 
  \usepackage{unicode-math}
  \defaultfontfeatures{Scale=MatchLowercase}
  \defaultfontfeatures[\rmfamily]{Ligatures=TeX,Scale=1}
\fi
\usepackage{lmodern}
\ifPDFTeX\else  
\fi
\IfFileExists{upquote.sty}{\usepackage{upquote}}{}
\IfFileExists{microtype.sty}{
  \usepackage[]{microtype}
  \UseMicrotypeSet[protrusion]{basicmath} 
}{}
\makeatletter
\@ifundefined{KOMAClassName}{
  \IfFileExists{parskip.sty}{%
    \usepackage{parskip}
  }{
    \setlength{\parindent}{0pt}
    \setlength{\parskip}{6pt plus 2pt minus 1pt}}
}{
  \KOMAoptions{parskip=half}}
\makeatother
\usepackage{xcolor}
\ifx\paragraph\undefined\else
  \let\oldparagraph\paragraph
  \renewcommand{\paragraph}[1]{\oldparagraph{#1}\mbox{}}
\fi
\ifx\subparagraph\undefined\else
  \let\oldsubparagraph\subparagraph
  \renewcommand{\subparagraph}[1]{\oldsubparagraph{#1}\mbox{}}
\fi

\providecommand{\tightlist}{%
  \setlength{\itemsep}{0pt}\setlength{\parskip}{0pt}}\usepackage{longtable,booktabs,array}
\usepackage{calc} 
\usepackage{etoolbox}
\makeatletter
\patchcmd\longtable{\par}{\if@noskipsec\mbox{}\fi\par}{}{}
\makeatother
\IfFileExists{footnotehyper.sty}{\usepackage{footnotehyper}}{\usepackage{footnote}}
\makesavenoteenv{longtable}
\usepackage{graphicx}
\makeatletter
\def\maxwidth{\ifdim\Gin@nat@width>\linewidth\linewidth\else\Gin@nat@width\fi}
\def\maxheight{\ifdim\Gin@nat@height>\textheight\textheight\else\Gin@nat@height\fi}
\makeatother
\setkeys{Gin}{width=\maxwidth,height=\maxheight,keepaspectratio}
\makeatletter
\def\fps@figure{htbp}
\makeatother
\NewDocumentCommand\citeproctext{}{}

\makeatletter
 \let\@cite@ofmt\@firstofone
 \def\@biblabel#1{}
 \def\@cite#1#2{{#1\if@tempswa , #2\fi}}
\makeatother
\newlength{\cslhangindent}
\newlength{\csllabelwidth}
\newenvironment{CSLReferences}[2] 
 {\begin{list}{}{%
  \setlength{\itemindent}{0pt}
  \setlength{\leftmargin}{0pt}
  \setlength{\parsep}{0pt}
  \ifodd #1
   \setlength{\leftmargin}{\cslhangindent}
   \setlength{\itemindent}{-1\cslhangindent}
  \fi
  \setlength{\itemsep}{#2\baselineskip}}}
 {\end{list}}
\usepackage{calc}

\KOMAoption{captions}{tableheading}
\makeatletter
\@ifpackageloaded{caption}{}{\usepackage{caption}}
\AtBeginDocument{%
\ifdefined\contentsname
  \renewcommand*\contentsname{Table of contents}
\else
  \newcommand\contentsname{Table of contents}
\fi
\ifdefined\listfigurename
  \renewcommand*\listfigurename{List of Figures}
\else
  \newcommand\listfigurename{List of Figures}
\fi
\ifdefined\listtablename
  \renewcommand*\listtablename{List of Tables}
\else
  \newcommand\listtablename{List of Tables}
\fi
\ifdefined\figurename
  \renewcommand*\figurename{Figure}
\else
  \newcommand\figurename{Figure}
\fi
\ifdefined\tablename
  \renewcommand*\tablename{Table}
\else
  \newcommand\tablename{Table}
\fi
}
\@ifpackageloaded{float}{}{\usepackage{float}}
\floatstyle{ruled}
\@ifundefined{c@chapter}{\newfloat{codelisting}{h}{lop}}{\newfloat{codelisting}{h}{lop}[chapter]}
\floatname{codelisting}{Listing}

\usepackage{amsthm}
\theoremstyle{plain}
\newtheorem{theorem}{Theorem}[section]
\theoremstyle{remark}
\AtBeginDocument{}

\makeatother
\makeatletter
\@ifpackageloaded{caption}{}{\usepackage{caption}}
\@ifpackageloaded{subcaption}{}{\usepackage{subcaption}}
\makeatother
\ifLuaTeX
  \usepackage{selnolig}  
\fi
\usepackage{bookmark}

\IfFileExists{xurl.sty}{\usepackage{xurl}}{} 
\hypersetup{
  pdftitle={Sampling Spiked Wishart Eigenvalues},
  pdfauthor={Thomas G. Brooks},
  colorlinks=true,
  linkcolor={blue},
  filecolor={Maroon},
  citecolor={Blue},
  urlcolor={Blue},
  pdfcreator={LaTeX via pandoc}}

\title{Sampling Spiked Wishart Eigenvalues}
\author{Thomas G. Brooks}
\date{}

\begin{document}
\maketitle
\begin{abstract}
Efficient schemes for sampling from the eigenvalues of the Wishart
distribution have recently been described for both the uncorrelated
central case (where the covariance matrix is \(\mathbf{I}\)) and the
spiked Wishart with a single spike (where the covariance matrix differs
from \(I\) in a single entry on the diagonal). Here, we generalize these
schemes to the spiked Wishart with an arbitrary number of spikes. This
approach also applies to the spiked pseudo-Wishart distribution. We
describe how to differentiate this procedure for the purposes of
stochastic gradient descent, allowing the fitting of the eigenvalue
distribution to some target distribution.
\end{abstract}

\renewcommand*\contentsname{Table of contents}
{
\hypersetup{linkcolor=}
\setcounter{tocdepth}{3}
\tableofcontents
}
\subsection{Introduction}\label{introduction}

The Wishart distribution is the distribution of random matrices of the
form \(\mathbf{W} = \mathbf{G} \mathbf{G}^T\) where \(\mathbf{G}\) is an
\(m \times n\) matrix with each column drawn iid from
\(N(0, \mathbf{\Sigma})\) for some \(m \times m\) covariance matrix
\(\mathbf{\Sigma}\), and \(\mathbf{G}^T\) denotes the transpose of
\(\mathbf{G}\). The most well-studied is the case where
\(\mathbf{\Sigma}\) is the identity matrix, which is sometimes referred
to as the uncorrelated central Wishart (Zanella, Chiani, and Win 2009).
A more general case is the spiked Wishart, where \(\mathbf{\Sigma}\) is
diagonal with \(k\) `spiked' eigenvalues (typically larger than 1) and
the remaining eigenvalues are 1. Here, \(k\) is usually small and \(m\)
large.

Recently, (Forrester 2024) described an efficient method to sample
eigenvalues of Wishart matrices in the cases of uncorrelated central
Wishart as well as for the spiked Wishart for the case \(k = 1\). Here,
we describe the generalization of that process to the general case
\(k\). This approach derives from the diagonalization strategy of
(Dumitriu and Edelman 2002) for the uncorrelated central case, which
carries over without modification to the \(k=1\) spiked Wishart case as
shown in (Forrester 2024). The diagonalization approach requires slight
modifications in the general \(k\) spiked Wishart distribution.
Moreover, we observe that this approach holds even in the pseudo-Wishart
case where the matrices \(W\) are not full rank, i.e., there are fewer
observations than variables (\(n < m\)).

By reparametrizing this sampling procedure, we are also able to compute
derivatives of the Wishart eigenvalues with respect to input spiked
values. This allows for the efficient use of stochastic gradient descent
for optimization problems involving the Wishart eigenvalue distribution.
In particular, we apply this to the problem of fitting the expectation
of the Wishart singular values to some empirical values.

\subsection{Preliminaries}\label{preliminaries}

We adopt the notation of (Dumitriu and Edelman 2002) when possible but
consider just the real case.

The straight-forward approach to sampling the eigenvalues of a Wishart
distribution, is to first sample \(\mathbf{G}\), compute
\(\mathbf{W} = \mathbf{G} \mathbf{G}^T\), and find its eigenvalues
(alternatively, one can compute the singular values of \(\mathbf{G}\)).
However, \(\mathbf{G}\) requires \(m \times n\) independent samples from
a normal distribution, and computing the eigenvalues of \(\mathbf{W}\)
(or the singular values of \(\mathbf{G}\)) is computationally demanding
for large \(m\) and \(n\). In (Dumitriu and Edelman 2002), it is instead
shown that a series of Householder reflections allow the transformation
of \(\mathbf{G}\) into a bidiagonal matrix with entries that are
independently distributed as the square roots of the \(\chi^2\)
distribution. This allows the sampling of just \(2 n\) random \(\chi^2\)
values and the orthogonal Householder reflections do not affect the
eigenvalues of \(\mathbf{G} \mathbf{G}^T\). This was used by (Forrester
2024) to sample the uncorrelated central Wishart or \(k=1\) spiked
Wishart.

A Householder reflection is a orthogonal matrix of the form
\(\mathbf{R} = \mathbf{I} - 2 vv^T\) for some unit vector \(v\). We use
the fact that for any vector \(w\) there is a Householder reflection
\(\mathbf{R}\) that takes \(w\) to \(\left\lVert{w}\right\rVert_2 e_1\)
where \(\left\lVert{w}\right\rVert_2\) is the 2-norm of the vector and
\(e_1 = [1, 0, \ldots, 0]\). If each entry \(w\) is iid
\(N(0, \sigma^2)\) then \(\left\lVert{w}\right\rVert_2\) is distributed
as \(\sigma \chi_k\) where \(k\) is the size of \(w\).

\subsection{\texorpdfstring{Sampling eigenvalues of the \(k\) spiked
Wishart}{Sampling eigenvalues of the k spiked Wishart}}\label{sampling-eigenvalues-of-the-k-spiked-wishart}

\begin{theorem}[]\protect\hypertarget{thm-sample}{}\label{thm-sample}

Let \(\mathbf{W} := \mathbf{G} \mathbf{G}^T\) where the entries of
\(\mathbf{G}\) are independent and
\((\mathbf{G})_{i,j} \sim N(0, \sigma_i^2)\) for all \(i,j\). Suppose
that \(\sigma_i = 1\) for all \(i > k\). Let
\(\mathbf{W}' := \mathbf{H} \mathbf{H}^T\) where \(\mathbf{H}\) has
independent entries satisfying:

\begin{enumerate}
\def\labelenumi{\arabic{enumi}.}
\tightlist
\item
  \((\mathbf{H})_{i,i}\) is \(\sigma_i \chi_{(n-i+1)}\) distributed,
\item
  \((\mathbf{H})_{i,i-k}\) is \(\sigma_i \chi_{(m-i+1)}\) distributed
  for \(i > k\),
\item
  \((\mathbf{H})_{i,j}\) is \(N(0,\sigma_i^2)\) for \(i-k < j < i\), and
\item
  \((\mathbf{H})_{i,j}\) is 0 for \(j > i\) or \(j < i-k\).
\end{enumerate}

Then the eigenvalues of \(\mathbf{W}\) and the eigenvalues of
\(\mathbf{W}'\) have the same distribution.

\end{theorem}

Here \(\sigma_i\) is the standard deviation of the \(i\)th variable,
with the first \(k\) being the `spike' eigenvalues and the remaining all
equaling \(1\). We denote \(N(0,\sigma^2)\) as the normal distribution
centered at 0 with variance \(\sigma^2\) and \(\chi_{\alpha}\) is the
square root of the \(\chi^2\) distribution with \(\alpha\) degrees of
freedom.

Each row \(\mathbf{G}_{i,\cdot}\) is a vector of multivariate normal
distribution with zero mean and covariance matrix
\(\sigma_i^2 \mathbf{I}_{n \times n}\). This distribution is
rotationally symmetric and so if \(\mathbf{R}\) is an orthogonal
\(n \times n\) matrix independent of a row \(\mathbf{G}_{i,\cdot}\),
then \((\mathbf{G} \mathbf{R})_{i,\cdot}\) has the same distribution as
\(\mathbf{G}_{i,\cdot}\).

Define \(\mathbf{G}_0 = \mathbf{G}\) and proceed inductively by choosing
two orthogonal matrices \(\mathbf{L}_i\) and \(\mathbf{R}_i\) and
defining \[
\mathbf{G}_i :=
  \begin{bmatrix} \mathbf{I} & 0  \\ 0 & \mathbf{L}_i \end{bmatrix}
  \mathbf{G}_{i-1}
  \begin{bmatrix} \mathbf{I}  & 0 \\ 0 & \mathbf{R}_i \end{bmatrix}.
\] First, we choose \(\mathbf{R}_i\) an
\((n - i + 1) \times (n - i +1)\) orthogonal matrix such that
\(x \mathbf{R}_i = \left\lVert{x}\right\rVert_2 e_1\) where and
\(x = [(\mathbf{G}_{i-1})_{i,i} \, (\mathbf{G}_{i-1})_{i,i+1} \, \ldots (\mathbf{G}_{i-1})_{i,n}]\).
This can be done with a Householder reflection and is a function only of
\(x\) the \(i\)th row of \(\mathbf{G}_{i-1}\). Therefore
\(\mathbf{R}_i\) is independent of every other of row of
\(\mathbf{G}_{i-1}\). Define
\[\mathbf{G}_{i-1}' = \mathbf{G}_{i-1} \begin{bmatrix} \mathbf{I} & 0 \\ 0 & \mathbf{R}_i \end{bmatrix}.\]

Second, we choose \(\mathbf{L}_i\) an \((m-i-k+1) \times (m-i-k+1)\)
orthogonal matrix such that
\(\mathbf{L}_i y^T = \left\lVert{y}\right\rVert_2 e_1^T\) where
\(y = [(\mathbf{G}_{i-1}')_{i+k,i}, \, (\mathbf{G}_{i-1}')_{i+k+1, i} \, \ldots (\mathbf{G}_{i-1}')_{m,i}]\).
In the case where \(i + k > m\), set \(\mathbf{L}_i\) to be the
\(0 \times 0\) matrix, making the left action on \(\mathbf{G}_{i-1}\)
the identity. Again, \(\mathbf{L}_i\) depends only upon the \(i\)th
column of \(\mathbf{G}'_{i-1}\) and so is independent of the rest of the
\(\mathbf{G}'_{i-1}\). Note that this is the bidiagonalization process
of (Dumitriu and Edelman 2002) except with \(\mathbf{L}_i\) restricted
to only ever act on the bottom \(m-k\) rows. Since in our case,
\(\sigma_i\) are not constant for \(i \leq k\), extending
\(\mathbf{L}_i\) into the first \(k\) rows would change the distribution
of all the entries.

Analogous to the bidiagonlization method, observe that \(\mathbf{G}_i\)
has the following properties for \(j \leq i\):

\begin{enumerate}
\def\labelenumi{\arabic{enumi}.}
\tightlist
\item
  \((\mathbf{G}_i)_{j,j}\) is \(\sigma_j \chi_{(n-j+1)}\) distributed,
\item
  \((\mathbf{G}_i)_{j+\ell,j}\) is \(N(0,\sigma_{j+\ell}^2)\) for
  \(\ell=1, \ldots, k-1\) distributed,
\item
  \((\mathbf{G}_i)_{j+k,j}\) is \(\sigma_{j+k} \chi_{(m-j-k+1)}\)
  distributed,
\item
  \((\mathbf{G}_i)_{j+\ell,j} = 0\) for \(\ell > k\) and
  \((\mathbf{G}_i)_{j,\ell} = 0\) for \(\ell > j\),
\item
  \((\mathbf{G}_i)_{\ell, \ell'}\) is \(N(0, \sigma_{\ell}^2)\) for
  \(\ell > i\) and \(\ell' > i\), and
\item
  all entries of \(\mathbf{G}_i\) are independent of each other.
\end{enumerate}

Therefore, \(\mathbf{H} := \mathbf{G}_n\) has the form as in
Theorem~\ref{thm-sample}, and it differs from \(\mathbf{G}\) only by a
series of orthogonal transformations. So the singular values of
\(\mathbf{G}_n\) are equal to the singular values of \(\mathbf{G}\) (and
to the square roots of the eigenvalues of
\(\mathbf{W}  = \mathbf{G} \mathbf{G}^T\)). This proves
Theorem~\ref{thm-sample}.

Moreover, \(\mathbf{H}\) is more efficient to sample than \(\mathbf{G}\)
and for small \(k\), its singular values can be computed efficiently
thanks to its sparse nature. Indeed, it is only non-zero in an
\((k+m) \times (k+m)\) submatrix, within which it has only
\(k+1\)-nonzero diagonals.

\subsection{Simulations}\label{simulations}

\begin{figure}

\centering{

\includegraphics{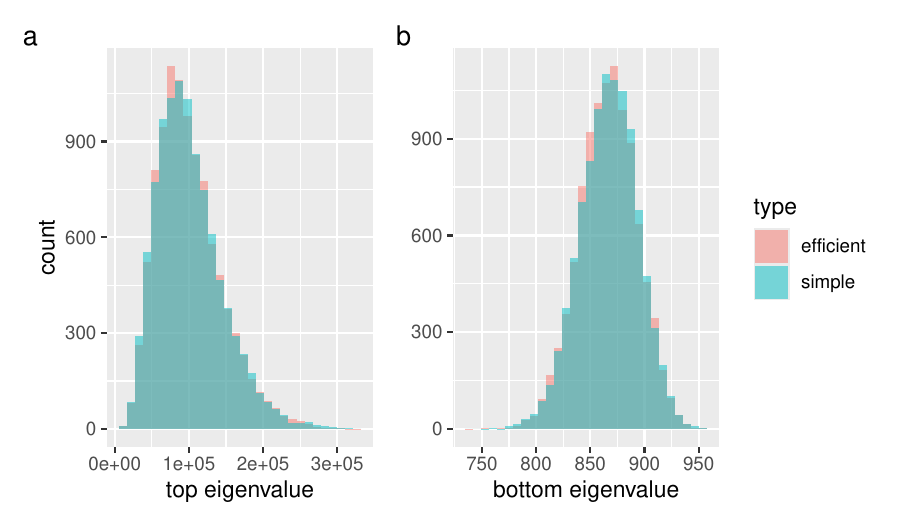}

}

\caption{\label{fig-histogram}Histograms of 10000 draws from each of the
simple (from G) and efficient (from H) methods. This has m=1000, n=1000
with spiked standard deviations 100, 30, 10. The simple method took
10.05s and the efficient method took 2.1s}

\end{figure}

We compare this method to a `simple' method where \(\mathbf{G}\) is
sampled directly from the multivariate normal distribution, see
Figure~\ref{fig-histogram}. With 10000 samples, the Kolmogorov-Smirnov
test comparing the two distributions of the top-eigenvalues has p=0.94
and the bottom-eigenvalues have p=0.06, failing to find any difference.

\begin{figure}

\centering{

\includegraphics{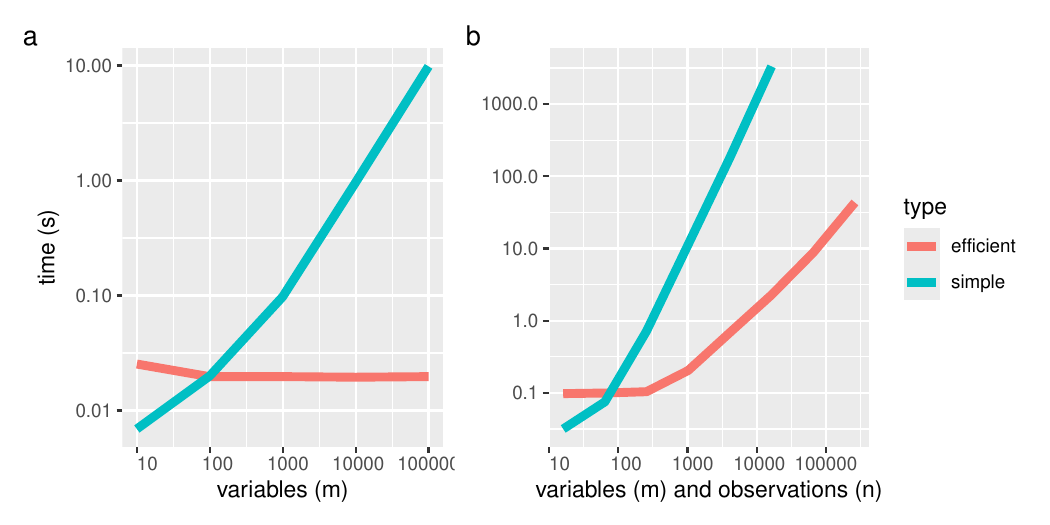}

}

\caption{\label{fig-scaling}Execution time required for 100 samples of
both methods. (a) Varying numbers of variables \(m\) while the number of
observations \(n\) is constant at 10. (b) Varying both \(m\) and \(n\)
together. In this case, sparse matrices are used and only the top 3
eigenvalues are computed. In both, the spiked standard deviations are
100, 30, 10.}

\end{figure}

To demonstrate efficiency, we consider scaling with \(m\), the number of
variables, see Figure~\ref{fig-scaling} a. Simple sampling of
\(\mathbf{G}\) directly outperforms the efficient sampling of
\(\mathbf{H}\) up to about 100 variables. However, the efficient
sampling has constant timing with the number of variables and remains
efficient as variable count increases. The constant time is due to the
constant value of \(n\), the number of observations, so that only a
constant number of rows and columns are non-zero. When instead varying
both \(m\) and \(n\) together, the efficient method again scales much
better than the simple method but is no longer close to constant time,
see Figure~\ref{fig-scaling} b. In this case, we use a sparse matrix SVD
computed for just 3 top singular values.

\subsection{Fitting and gradient
descent}\label{fitting-and-gradient-descent}

Consider the problem of identifying which spiked eigenvalues to use
based off desired distribution of eigenvalues of the Wishart matrix.
Since we have no analytic solution for this distribution (outside of the
\(k=1\) case (Zanella and Chiani 2020)), we instead use stochastic
gradient descent on random samples from this distribution. In this
section, we operate on the singular values of \(\mathbf{H}\), which are
the square roots of the eigenvalues of \(\mathbf{W}\), for simplicity.
To do so, we need to differentiate the random sample with respect to the
spiked values. This becomes an application of the well-known
reparametrization trick from machine learning, introduced for
variational autoencoders (Kingma and Welling 2013). This trick is to
separate out the random sampling and consider it to be fixed while
taking gradients. In particular, notice that
\[\mathbf{H}' := \begin{bmatrix} \sigma_1^{-1} & & \\ & \ddots & \\ & & \sigma_m^{-1} \end{bmatrix} \mathbf{H}\]
has distribution that does not depend upon
\(\sigma_1, \ldots, \sigma_m\). Indeed, each entry in \(\mathbf{H}'\) is
either distributed as standard normal or \(\chi\) with a degrees of
freedom not depending upon \(\sigma_r\) for any \(r\). Therefore, we
consider the gradient with respect to the \(\sigma_r\) of
\(\mathbf{H}'\) to be zero. Moreover,
\begin{align*}
\frac{\partial \mathbf{H}_{i,j}}{\partial \sigma_{r}}
  &= \begin{cases} \mathbf{H}'_{r,j}\hphantom{\sigma_r^{-1}} & r = i \\ 0 & \mbox{otherwise} \end{cases} \\
  &= \begin{cases} \sigma_{r}^{-1} \mathbf{H}_{r,j} & r = i \\ 0 & \mbox{otherwise} \end{cases}
.
\end{align*}

Next, we consider the singular value decomposition
\(\mathbf{H} = \mathbf{U} \mathbf{D} \mathbf{V}^T\) where \(U\) and
\(V\) are orthogonal matrices and \(D\) is diagonal containing the
singular values along its diagonal. It is well-known that the partial
derivative of \(d_\ell\), the \(\ell\)th singular value, with respect to
the entry \(\mathbf{H}_{i,j}\) is
\(\mathbf{U}_{i,\ell} \mathbf{V}_{j,\ell}\). Then
\begin{align*}
  \frac {\partial d_\ell} {\partial \sigma_r}
  = \sum_{i,j} \frac{\partial d_\ell} {\partial \mathbf{H}_{i,j}} \frac{\partial \mathbf{H}_{i,j}} { \partial \sigma_r }
  = \sum_{j=1}^n \mathbf{U}_{r,\ell} \mathbf{V}_{j,\ell} \mathbf{H}'_{r,j}
  = \sigma_r^{-1} \mathbf{U}_{r,\ell} \sum_{j=1}^n \mathbf{H}_{r,j} \mathbf{V}_{j,\ell}
\end{align*}
for \(r \leq k\). This gives the derivative of a single sample. To
extend this to an estimate of the derivative of the expected value, we
can simply take the mean over a large sample of \(\mathbf{H}\) matrices.

\begin{figure}

\centering{

\includegraphics{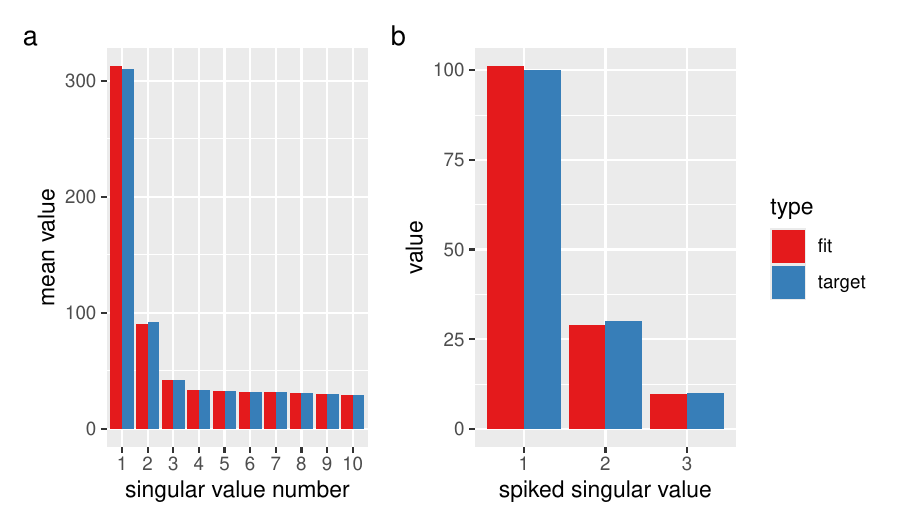}

}

\caption{\label{fig-fitting}Demonstration of fitting spiked singular
values to give desired output singular values of \(\mathbf{H}\). Using
the same values as Figure 1 as the target, we optimized three spiked
singular values to give the best match of the expectation of the
eigenvalues to the target ones. We plot (a) the means of singular values
of \(\mathbf{H}\) and (b) the spiked singular values in target and fit
distributions.}

\end{figure}

As a demonstration, we implement a Levenberg-Marquardt minimizer of the
sum of squared errors for the singular values of \(\mathbf{H}\) to some
target specified singular values. Using the same values as in
Figure~\ref{fig-histogram} as the true target values, we computed the
mean singular values and fit three spiked singular values to have these
as the expected values. These were then used to sample the mean of the
singular values, providing a good match, see Figure~\ref{fig-fitting} a.
Moreover, the fit spiked singular values were themselves a close match
to the original singular values, see Figure~\ref{fig-fitting} b.

\subsection{Discussion}\label{discussion}

The case where \(m > n\) is of particular interest to the study of
`omics' data sets, such as genomics, transcriptomics, proteomics and
others. These data are costly to collect and assay, so typically few
samples are available, but each sample has a deep wealth of information.
One goal in this realm is the accurate simulation of data sets, for
purposes including the benchmarking and optimization of the accuracy of
computational pipelines. The spiked Wishart is a natural use-case for
this situation, since the low number of samples means that a complete
covariance matrix cannot be directly estimated. For example, the
\texttt{corpcor} software package (Schäfer and Strimmer 2005)
(Opgen-Rhein and Strimmer 2007) performs estimation of a correlation
matrix as a linear combination of a sample covariance matrix (which is
low-rank when \(m > n\)) and the identity matrix, just as a spiked
Wishart does. Instead, we may wish to simulate data based off a real
data set. In that case, we wish to simulate data via a spiked Wishart
distribution and to have it match (in expectation) the eigenvalues of
the covariance matrix of the real data set. This sampling approach
allows fitting the eigenvalue distribution to such data.

We provide in Supplementary Material an implementation of this sampling
method in the R programming language (using the \texttt{Matrix} and
\texttt{sparsesvd} packages for sparse matrix computations) and for
comparison the simple (`brute-force') method of sampling \(\mathbf{G}\)
directly for comparison. This also includes routines for computing
Jacobians of the sampled singular values and performing fitting of
expected values of the Wishart eigenvalues. Lastly, it contains the
source code for this document and all of its figures.

\subsection*{References}\label{references}
\addcontentsline{toc}{subsection}{References}

\phantomsection\label{refs}

\end{document}